# Scaling relations for parallel disc Capacitance with Dielectric media

C. V. Krishnamurthy[†], Gokul Raj R[*]

Measurement and Modeling Lab

Department of Physics, Indian Institute of Technology Madras

Chennai-600036, India.

**Abstract**

Total capacitance of a dielectric filled circular disc configuration (radius $r$ and separation $d$) has been numerically evaluated over a wide range of relative permittivities ($1 \leq \varepsilon_r \leq 80$) and aspect ratios ($0.001 \leq d/r \leq 10$). A new perspective is presented by which, the total capacitance, for a given disc radius $r$, is found to depend only on a simple scaled disc separation, $d^* = d/\varepsilon_r$ for the whole range of aspect ratios and relative permittivities considered. The scaling notion stems from noting that the path length has to be larger than what it would be in free space to have the same potential difference across the charged discs since the electric field strength is reduced within a homogenous dielectric medium by $\varepsilon_r$. This scaling permits translation of practical measurements across a range of aspect ratios and relative permittivities. Further, the scaling feature allows for any computational approach that determines the total capacitance in free space ($\varepsilon_r = 1$), to be applicable for a range of dielectric media having relative permittivities $1 \leq \varepsilon_r \leq 80$. It is believed that the scaling would become better with increasing relative permittivities for practically relevant separations.

**Introduction**

The capacitance for identical thin circular conducting discs, charged equally and oppositely, with a dielectric medium arranged in the configuration shown in Fig. 2(b) can be expressed in terms of charge $Q$ on an electrode and potential difference $V$ between the discs as

$$C = \frac{Q}{V} = \frac{\oint \vec{D}.\overrightarrow{dA}}{\int_0^d \vec{E}.\overrightarrow{dl}} \tag{1}$$

Appendix I contains the definitions for the various quantities introduced here and the background associated with equation (1). The charge $Q$ on a disc is obtained through Gauss's law by integrating the outward flux (for positive charge) of the displacement field over any surface that encloses the disc chosen conveniently for purposes of evaluation. An integration over the other disc would lead to the total charge on that disc but of opposite sign. The potential difference $V$ is essentially the work done in transferring a unit charge from one disc to the other and is path-independent. It is obtained by carrying out the line integral of the electric field over any path connecting the two discs chosen conveniently for purposes of evaluation. The shortest paths are made up of segments locally perpendicular to the succession of equipotential surfaces between the two discs.

When these two oppositely charged discs are brought close to each other in free space, there is a redistribution of the charges on each disc due to induction, the extent of which depends on the separation between the discs. At very large separations, induction effects are small and the charge distribution on a disc would tend to that of an isolated charged disc for which charge is distributed over both the sides of the disc symmetrically. It is well known that the charge distribution, $\sigma_{nu}(\rho)$, *on each side* of an isolated disc-like conductor is highly non-uniform and is given by MacDonald[12] as,

$$\sigma_{nu}(\rho) = \frac{1}{2}Q \frac{1}{2\pi r\sqrt{r^2-\rho^2}} \qquad (2)$$

where, $r$ is the radius of the circular disc and $\rho$, the radial variable, lies in the range ($0 \leq \rho \leq r$). The given total charge $Q$ distributes itself equally on *both* sides of the isolated disc as shown in Fig. 1, such that the density is highest at the edges and falls rapidly to a finite value at the disc center with a very small region around the center where the charge density can be said to be uniform.

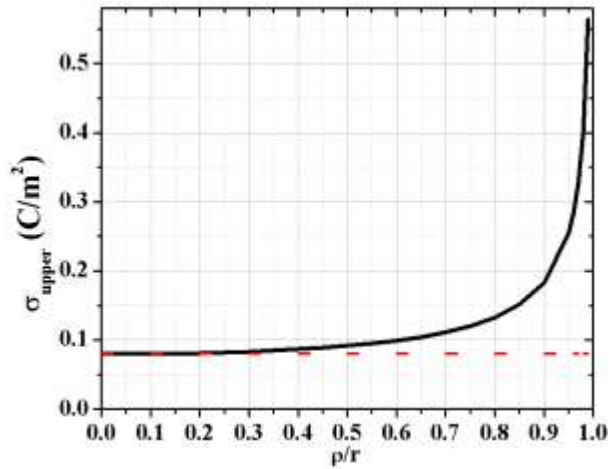

Fig. 1. Non-uniform radial charge distribution on the upper face of a conducting circular disk of unit radius (carrying one unit of charge). The dotted red curve represents the uniform charge density (= $1/4\pi$ =0.0795).

For very small disc separations, induction effects are strong due to which considerable redistribution of the charges occurs, leading to asymmetric charge distributions on each disc. Specifically, for a parallel disc configuration, the surfaces of each disc facing each other tend to have greater charge than their respective back surfaces that face away from each other. An idealization would be to consider *all* charge on each disc to exist only on their inner surfaces (so that, $\vec{E}_{in} \neq 0$, and $\vec{E}_{out} = 0, \vec{E}_{sides} \approx 0$). If further, the charge is taken to be *uniformly* distributed across each of the discs' inner surfaces, the surface and line integrals can be evaluated easily. With $\vec{E}_{in} \neq 0$, $\vec{D}_{in} = \varepsilon_0 \vec{E}_{in}$, where $\varepsilon_0$ is the free space permittivity, and with $\vec{E}_{out} = 0$, $\vec{E}_{sides} = 0$, the surface integral yields $\varepsilon_0 |\vec{E}_{in}| A$. Here, $A$ is the area of the inner surface of a disc. As the field lines are taken to be uniform in the region between the discs, the line integral results in $|\vec{E}_{in}| d$, where $d$ is the separation between the parallel discs. These two simplifications lead to the well-known result,

$$C_{ideal} = \varepsilon_0 A / d \qquad (3)$$

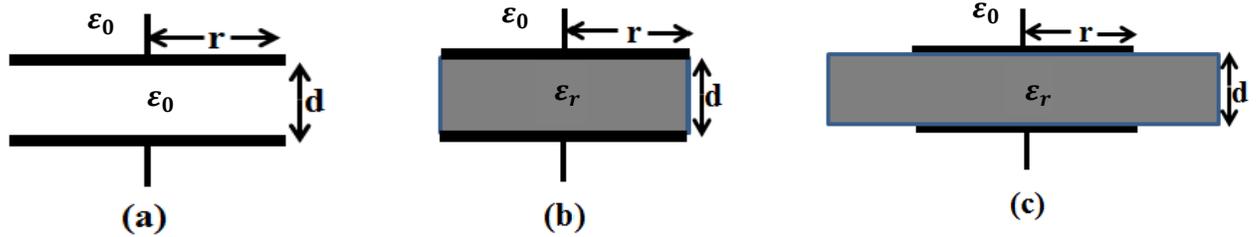

Fig.2. Various configurations of identical circular disk parallel electrode capacitor (a) same dielectric in every region (b) dielectric extending up to the geometric limit (c) dielectric extending beyond the geometric limits.

However, charges do get concentrated at the edges of the discs and do get distributed on the back sides of the discs as well, leading to non-uniform charge distributions and non-uniform fields at the edges and extending to the back side of each disc.

The problem of computing the total capacitance when the actual non-uniform charge distribution is considered, and when the associated non-uniform fields (termed hereinafter as fringe fields) are taken into account, has been extensively investigated analytically, semi-analytically and numerically.[2,3,5,9,10,11]

Carlson and Illman[3] employ Fourier cosine series to evaluate the final integral equation for the total capacitance numerically over a wide range of aspect ratios that were not dealt with until then. The total capacitance in free space obtained by Carlson and Illman[3] is shown in Fig. 3.

The problem of a dielectric loaded capacitor has received much less attention. Three geometries considered in the literature are shown in Figure 2(a)[3,9,10,13,14], Figure 2(b)[5,6] and Figure 2(c)[1,6,11]. A few attempts have been made to evaluate the capacitance of a dielectric loaded capacitor as a function of geometric spacing between the electrodes as well as a function of the dielectric constant[1,5,6,11] although the geometries are not the same. The challenge posed by the presence of the dielectric medium is two-fold: the change in the fringe field contributions to the total capacitance and the extraction of the dielectric constant from the total capacitance.

In the present work, the results of numerical (FEM) calculations[4] of total capacitance for a wide range of aspect ratios and dielectric constants obtained recently[5] are examined for possible scaling relations that helps understand the role of the dielectric media and fringe fields in determining the total capacitance as a function of the aspect ratio. Sections I and II describe the features of capacitance in free space and in the presence of dielectric media respectively. Section III explores two scaling relations and it is shown that it is possible to map the dielectric-loaded total capacitance onto the free space total capacitance in terms of a scaled electrode separation over a wide range of aspect ratios and a wide range of relative permittivities. Section IV presents the conclusions of the current investigations.

**I. Total Capacitance when $\varepsilon_r = 1$**

Figure 2 presents the total capacitance results for $\varepsilon_r = 1$ using normalized values. The total capacitance has been normalized by $4\varepsilon_0 r$, which is the parallel combination of self-capacitances of two isolated circular thin electrodes of radius $r$. The electrode spacing is expressed in terms of aspect ratio $d/r$.

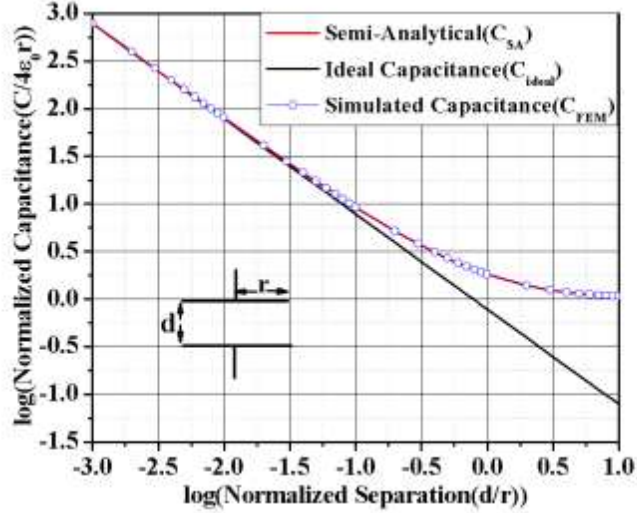

Fig. 3. Comparison between semi-analytical and thin electrode numerical (FEM) results for symmetric circular electrode configuration

Figure 3 highlights two features: (a) the "ideal" capacitance given by $C_{ideal} = \varepsilon_0 A/d = \varepsilon_0 \pi r^2/d$ leads to a straight line in the log-log scale for the entire range of aspect ratios chosen, and (b) the total capacitance increases systematically, as aspect ratio increases, with respect to the ideal capacitance. However, Fig. 3 also indicates that the range of aspect ratios ($0.001 \leq d/r < 0.1$) for which the deviation due to fringe field contributions are small and may be either ignored or handled using Kirchoff correction.[7, 8]

The systematic increase of the total capacitance (at constant potential difference) from the 'ideal' capacitance as the aspect ratio increases is due to the ability of the discs to accommodate more charges that spread towards the edges and on the back surface leading to non-uniform charge distribution which in turn sets up a non-uniform field distribution. This can also be viewed as the weakening of the induction effects of one charged electrode on the other to hold *all* the charges on the inner surfaces at larger aspect ratios.

The thin electrode FEM calculations[5] can be seen to agree with the semi-analytical results[3]. It has been found that the agreement is up to the first significant digit in the pico-farad scale in the total capacitance. Thicker electrodes lead to distinct and measurable results as has been reported recently[5].

## II. Total Capacitance when $\varepsilon_r > 1$

The total capacitance for a medium sandwiched between symmetric circular discs can, in general, be a function of $r$, $d$, $\varepsilon_r$. For small separations or small aspect ratios, the total capacitance can be well approximated by $\varepsilon_r \varepsilon_0 A/d$ since the displacement field inside the region between the discs can be expressed as, $\vec{D}_{in} = \varepsilon_0 \varepsilon_r \vec{E}_{in}$ and assuming as before that charges reside only on the inner surfaces of both the discs. This is essentially the scaling of the ideal capacitance by $\varepsilon_r$, i.e., $C = \varepsilon_r C_{ideal}$.

Figure 4(a) shows the FEM results for relative permittivities in the range $1 \leq \varepsilon_r \leq 80$ over the same range of values for the aspect ratio $(d/r)$ for a configuration shown in Fig.2(b). With increasing value of the relative permittivity, two features can be noted from Fig. 4(a): (i) the total capacitance have seen to increase systematically as the relative permittivity increases, and (ii) the total capacitance as a function of the aspect ratio tends to straighten, in the log-log scale, more and more as the relative permittivity increases.

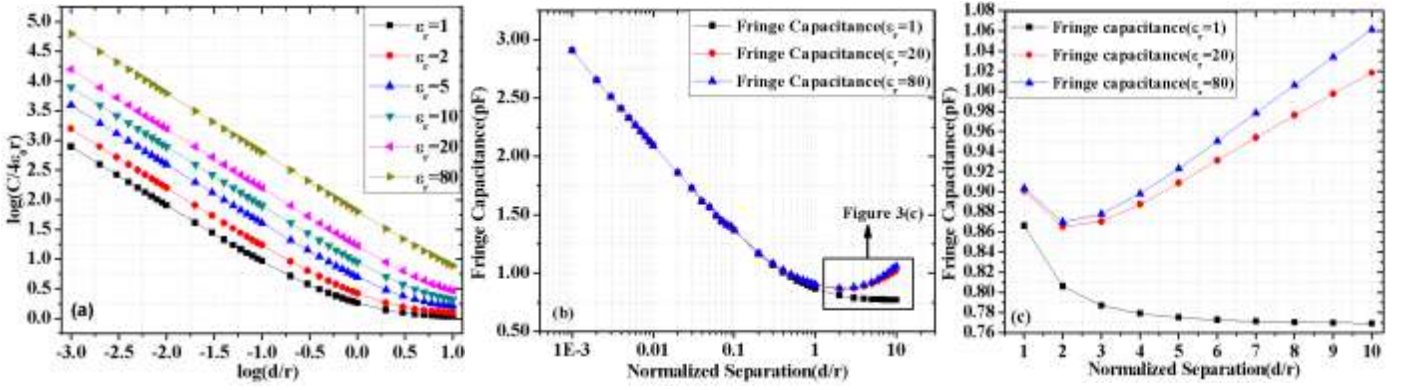

Fig. 4. (a) Variation of normalized capacitance with normalized separation for various relative permittivities in log-log scale. (b) Variation of fringe capacitance for various dielectrics and aspect ratios. (c) Inset in (b) expanded and employing linear scales.

The fringe field contribution to the total capacitance can be defined by the following relation $C_{fringe} = C_{total} - \varepsilon_r C_{ideal}$ where $\varepsilon_r$ is the relative permittivity of the medium between the electrodes and $C_{ideal} = \varepsilon_0 A/d$ is the geometrical capacitance. It is seen from Fig. 4(b) that the fringe capacitance is almost independent of the relative permittivity in the range $(0.001 \leq d/r < 0.1)$, and weakly dependent on the relative permittivity in the range $(0.1 < d/r < 1.0)$. Beyond $d/r > 1.0$, there is a strong distinction between $\varepsilon_r = 1$ case and $\varepsilon_r > 1$ cases, although there is very little change between $\varepsilon_r = 20$ and $\varepsilon_r = 80$ cases as can be seen from Fig.4(c). As the aspect ratio increases, when the medium between the discs is different from that of the free space outside, the charges tend to go towards the edges and onto the back surface leading to increasing fringe fields that results in the upward trends seen in Fig.4(c). In other words, the dielectric is not able to 'hold' all the charges on the inner surfaces of the discs at larger aspect ratios. However, as all these changes are less than a picofarad, the total capacitance tends to behave more like $\varepsilon_r C_{ideal}$ and appears to straighten in Fig. 4(a) for larger aspect ratios and for higher relative permittivities. It is important to note that (a) $C_{fringe}$ is found not to scale with $\varepsilon_r$, and (b) there is no closed form expression for the semi-analytic result in Fig. 3 that can describe the total capacitance for larger aspect ratios, even for $\varepsilon_r = 1$.

**Section III**

It is interesting to seek scaling relations for the total capacitance that is valid for larger separations when $\varepsilon_r > 1$ using the semi-analytical result for $\varepsilon_r = 1$. Two such attempts are described in what follows:

A simple scaling for the total capacitance, valid for small aspect ratios, when $\varepsilon_r > 1$, is given by $C = \varepsilon_r C_{ideal}$. To test this scaling idea over the whole range of aspect ratios and relative permittivities, the numerically evaluated total capacitance is normalized by $4\varepsilon_0\varepsilon_r r$, following what was done earlier ($4\varepsilon_0 r$) but now scaled by $\varepsilon_r$ and plotted as function of $d/r$ as shown in Figure 5.

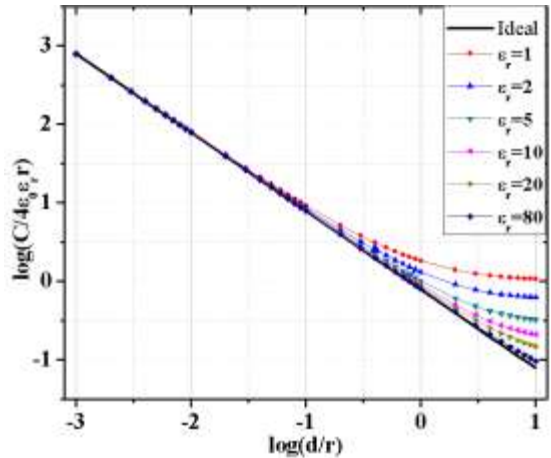

Fig. 5. Variation of normalized capacitance with normalized separation for various dielectric constants.

The trends in Fig. 5 show that the ideal capacitance neglecting fringe fields, and the total capacitance accounting for the fringe fields when $\varepsilon_r = 1$, set the boundaries, as it were, for all the cases when $\varepsilon_r > 1$. It is interesting to note from Fig. 5 that as the relative permittivity increases, the behavior tends to straighten (in the log-log scale) and approach the ideal capacitance behavior. This can be traced to the trends in Fig. 4(b) where fringe field contributions are found to be nearly independent of relative permittivity over the range ($0.001 \leq$ $d/r < 0.1$). This may be understood as follows: media with high relative permittivities 'hold' the charges on the inner surfaces of the discs and confine the field flux within the region occupied by the discs justifying the approximation leading to $C = \varepsilon_r \varepsilon_0 A/d$. Beyond $d/r > 0.1$, the media with high relative permittivities tend to keep the fringe field contributions small but non-negligible. From a practical point of view, the trends for $\varepsilon_r > 1$ indicate that the larger the value of $\varepsilon_r$, the greater is the range in electrode separation over which the total capacitance can be approximated well by a simple scaling of the ideal capacitance by $\varepsilon_r$.

Alternately, the total capacitance, for small aspect ratios, can be re-expressed as $C = \varepsilon_0 A/(d/\varepsilon_r) = \varepsilon_0 A/d^*$. The electrode separation $d$ is now scaled to $d^*$, where, $d^* = d/\varepsilon_r$. Thus the total capacitance, for a given electrode separation $d$, and with a dielectric ($\varepsilon_r > 1$) is the same as the total capacitance in free space, but for an equivalent electrode separation, $d^*$. The basis for the scaling of $d$ is as follows: From $\oint \vec{D} \cdot \vec{dA} = Q$, and the constitutive relation for isotropic dielectric media, $\vec{D} = \varepsilon_0 \varepsilon_r \vec{E}$, an expression for $|\vec{E}|$ is obtained, which for small aspect ratios can be approximated to give $|\vec{E}| = Q/\varepsilon_0 \varepsilon_r A$, where $A$ is the area of the inside surface of the disc. It can be seen that the electric field strength in the presence of a dielectric is reduced by $\varepsilon_r$ from that in free space (given by $Q/\varepsilon_0 A$). The potential difference $V$, defined by $\int_0^d \vec{E} \cdot \vec{dl}$, is evaluated, for small aspect ratios, using the above result for the electric field strength $|\vec{E}| = Q/\varepsilon_0 \varepsilon_r A$ giving the following result: $V = (Q/\varepsilon_0 \varepsilon_r A) d$. Since the electric field strength in the presence of a dielectric medium is reduced, the path length has to be $d$ to obtain the same potential difference $V$ as that in free space obtained using a shorter path length $d^*$ which is given by $V = (Q/\varepsilon_0 A) d^*$. The capacitance in free space turns out to be $\varepsilon_0 A/d^*$.

To test this scaling idea over the whole range of aspect ratios and relative permittivities, Figure 6 has been generated by normalizing the numerically evaluated total capacitance by $4\varepsilon_0 r$ and by plotting this with respect to $d^*/r$. Significantly, the validity of this scaling is observed to hold over the whole range of aspect ratios ($0.001 \leq d/r < 10$) and relative permittivities considered ($1 \leq \varepsilon_r \leq 80$).

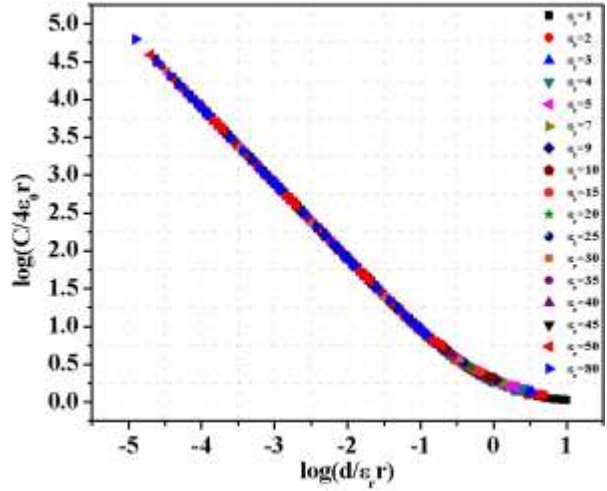

Fig. 6. Scaled Capacitance for a wide range of dielectrics ($1 \leq \varepsilon_r \leq 80$) and aspect ratios ($0.001 \leq d/r < 10$).

The notion that total capacitance is scaled by the relative permittivity is widely employed but is of limited applicability as seen from the previous discussion. A newer perspective, of much wider applicability, emerges from the notion that the total capacitance with a dielectric ($\varepsilon_r$) and electrode spacing $d$, is the same as the total capacitance in free space with an equivalent electrode spacing, $d^* = d/\varepsilon_r$. This scaling behavior is non-trivial since it includes fringe field contributions as well and does not involve any approximations or assumptions.

The scaling of the total capacitance has significant practical consequences in estimating the relative permittivity of unknown materials. All one needs to know is the total capacitance for free space as a function of the aspect ratio (either in a graphical form or as a table). Using the measured total capacitance, the scaled disc separation ($d^*$) is "read off" from the graph or table and is compared against the physical separation ($d$) to obtain the dielectric constant of the unknown material (assuming disc radius to be a constant) without any restrictions on sample thickness.

**IV. Conclusions**

Total capacitance of a dielectric filled circular disc configuration has been numerically evaluated over a wide range of relative permittivities ($1 \leq \varepsilon_r \leq 80$) and aspect ratios ($0.001 \leq d/r \leq 10$). The total capacitance has been found to depend only on a simple scaled separation over the whole range of relative permittivities and the whole range of aspect ratios examined in the present study. While it is known that the electric field strength in a homogeneous dielectric medium is reduced by $\varepsilon_r$ giving rise to the present scaling notion for the path length, what has been shown in the present study is that this simple scaling notion appears to work not only for small aspect ratios but for larger aspect ratios as well over a wide range of relative permittivities. It is expected that this scaling behavior may also be satisfied for the configuration in which layered media occupy the space between the two parallel conducting discs.

Through the scaling notion, the problem of total capacitance of a parallel disc configuration with a dielectric is reduced to the problem of total capacitance of a parallel disc configuration in free space. The intuitive yet non-trivial scaling relation found from the present investigation allows any computational approach that determines the total capacitance for $\varepsilon_r = 1$, to be applicable for any dielectric loaded configuration. The medium has been assumed not to extend beyond the disc diameter in the entire study.

**Appendix I**

In electrostatics involving charges on conductors and polarized dielectric media, as in the present context of a dielectric medium placed between oppositely charged parallel conducting discs, the following relations, $\vec{\nabla} \cdot \vec{E} = \rho_{total}/\varepsilon_0$ and $\vec{\nabla} \times \vec{E} = 0$ are satisfied for the electric field. The total charge density $\rho_{total}$ is made up of the free charge density $\rho_{free}$ on conductors and the bound charge density $\rho_{bound}$ that appears on the surface of the polarized dielectric medium. The response of the dielectric medium to external fields is described by $\vec{P}$, the induced dipole moment per unit volume, which satisfies the relation $\vec{\nabla} \cdot \vec{P} = -\rho_{bound}$. The negative sign reflects the oppositely charged induced response from the medium. The contributions from free charges and bound charges may be distinguished by using the displacement field $\vec{D}$ defined as $\vec{D} = \varepsilon_0 \vec{E} + \vec{P}$ which satisfies the relation, $\vec{\nabla} \cdot \vec{D} = \rho_{free}$. When the medium is isotropic and the field strength is small (compared to internal fields), the induced dipole moment per unit volume is expected to be proportional to the inducing electric field and is expressed as $\vec{P} = \chi \varepsilon_0 \vec{E}$ where $\chi$ is termed the susceptibility of the dielectric medium. The relation $\vec{D} = \varepsilon_0 \vec{E} + \vec{P}$ may be expressed as $\vec{D} = \varepsilon \vec{E} = \varepsilon_r \varepsilon_0 \vec{E}$ in terms of permittivity $\varepsilon$ of the medium or in terms of the relative permittivity $\varepsilon_r$ of the medium with respect to the permittivity free space $\varepsilon_0$.

The integral equivalents to the differential forms above are $\oint \vec{E} \cdot \vec{dl} = 0$ and $\oint \vec{D} \cdot \vec{dA} = \int \rho_{free} dV = Q$ respectively. $Q$, in the present context, is the charge on the conductor when the integral is taken over an imaginary surface enclosing the conductor.

The capacitance $C$ of a pair of conductors in a specific geometric configuration in free space is defined as the charge $Q_{fs}$ that can be put on the conductors by a source such that the potential difference between the two conductors is $V$ and is defined as

$$C = Q_{fs}/V$$

When a dielectric medium is introduced between the two charged conductors, as in the parallel disc configuration under study, the medium gets polarized and a bound charge appears on the surface of the medium. The source compensates for the reduction of the amount of free charge on the conductors, due to the induced bound charge of the dielectric medium, thereby increasing the amount of charge on the conductors to $Q$ ($> Q_{fs}$) at the same potential difference. The capacitance is said to have been increased due to the introduction of the dielectric medium. The increase in capacitance depends on the relative permittivity of the dielectric medium, more the relative permittivity, more the increase in the capacitance. It is important to note that the spacing between the charged conductors is not altered when the dielectric medium is introduced.

[†]Electronic mail : cvkm@iitm.ac.in
[*] Electronic mail: gokulraj@physics.iitm.ac.in,